\documentclass[prb,reprint,twocolumn,superscriptaddress,noshowpacs,notitlepage,longbibliography,10pt,citeautoscript]{revtex4-2}%
\usepackage{graphicx,bm,times}
\graphicspath{ {./figures/main/}{./figures/supp/}}
\usepackage{booktabs}
\usepackage{lipsum}
\usepackage{amsmath}
\usepackage{amsfonts}
\usepackage{amssymb}
\usepackage{mathtools}
\usepackage{color}
\usepackage{hyperref}
\usepackage{chemformula}
\usepackage{acro}
\hypersetup{colorlinks=true,allcolors={blue}}
\DeclareAcronym{VASP}{short = VASP, long = Vienna Ab initio Simulation Package}
\DeclareAcronym{AFM}{short = AFM, long = anti-ferromagnetic}
\DeclareAcronym{LEED}{short = LEED, long = low energy electron diffraction}
\DeclareAcronym{ARPES}{short = ARPES, long = angle-resolved photoemission spectroscopy}
\DeclareAcronym{FS}{short = FS, long = Fermi surface}
\DeclareAcronym{DFT}{short = DFT, long = density functional theory}
\DeclareAcronym{SOC}{short = SOC, long = spin-orbit coupling}
\DeclareAcronym{VHs}{short = VHs, long = Van Hove singularity, short-plural-form = VHss, long-plural-form = Van Hove singularities}
\DeclareAcronym{STM}{short = STM, , short-plural-form = STM, long = scanning tunneling microscopy, long-plural-form = scanning tunneling microscope}
\DeclareAcronym{STS}{short = STS, long = scanning tunneling spectroscopy}
\DeclareAcronym{BZ}{short = BZ, long = Brillouin zone}
\DeclareAcronym{RP}{short = RP, long = Ruddlesden-Popper}
\DeclareAcronym{FM}{short = FM, long = Ferromagnetic}
\DeclareAcronym{DOS}{short = DOS, long = density of states}
\DeclareAcronym{MIT}{short = MIT, long = metal-to-insulator transition}
\DeclareAcronym{UHV}{short = UHV, long = ultra-high vacuum}

\begin{document}
%%%%%%%%%%%%%%%%%%

%%%%%%%%%%%%%%%%%%%%%%%%%%%%%%%%%%%%%%%%%%%%%%%%%%%%%%%%%%%%%%%%%%%%%%%%%%%%%%
\title{A correlated insulator at the surface of the polar metal \ch{Ca3Ru2O7}}
%%%%%%%%%%%%%%%%%%%%%%%%%%%%%%%%%%%%%%%%%%%%%%%%%%%%%%%%%%%%%%%%%%%%%%%%%%%%%%
\author{Daniel Halliday}
\affiliation{SUPA, School of Physics and Astronomy, University of St Andrews, St Andrews KY16 9SS, UK} 
\affiliation{Diamond Light Source Ltd, Diamond House, Harwell Science and Innovation Campus, Didcot, OX11 0DE, UK}
\author{Izidor Benedi\v{c}i\v{c}}
\affiliation{SUPA, School of Physics and Astronomy, University of St Andrews, St Andrews KY16 9SS, UK}
\affiliation{Max Planck Institute for Chemical Physics of Solids, N{\"o}thnitzer Strasse 40, 01187 Dresden, Germany}
\author{Andela Zivanovic}
\affiliation{SUPA, School of Physics and Astronomy, University of St Andrews, St Andrews KY16 9SS, UK}
\affiliation{Max Planck Institute for Chemical Physics of Solids, N{\"o}thnitzer Strasse 40, 01187 Dresden, Germany}
\author{Masahiro Naritsuka}
\affiliation{SUPA, School of Physics and Astronomy, University of St Andrews, St Andrews KY16 9SS, UK}
\author{Brendan Edwards}
\affiliation{SUPA, School of Physics and Astronomy, University of St Andrews, St Andrews KY16 9SS, UK}
\author{Tommaso Antonelli}
\affiliation{SUPA, School of Physics and Astronomy, University of St Andrews, St Andrews KY16 9SS, UK}
\author{Naoki Kikugawa}
\affiliation{National Institute for Materials Science, Tsukuba, Ibaraki 305-0003, Japan}
\author{Dmitry A. Sokolov}
\affiliation{Max Planck Institute for Chemical Physics of Solids, N{\"o}thnitzer Strasse 40, 01187 Dresden, Germany}
\author{Craig~Polley}
\affiliation{MAX IV Laboratory, Lund University, 221 00, Lund, Sweden}
\author{Andrew P. Mackenzie}
\affiliation{SUPA, School of Physics and Astronomy, University of St Andrews, St Andrews KY16 9SS, UK}
\affiliation{Max Planck Institute for Chemical Physics of Solids, N{\"o}thnitzer Strasse 40, 01187 Dresden, Germany}
\author{Georg Held}
\affiliation{Diamond Light Source Ltd, Diamond House, Harwell Science and Innovation Campus, Didcot, OX11 0DE, UK}
\author{Phil D. C. King}\email{pdk6@st-andrews.ac.uk}
\affiliation{SUPA, School of Physics and Astronomy, University of St Andrews, St Andrews KY16 9SS, UK}
\author{Peter Wahl}\email{wahl@st-andrews.ac.uk}
\affiliation{SUPA, School of Physics and Astronomy, University of St Andrews, St Andrews KY16 9SS, UK}
\affiliation{Physikalisches Institut, Universit{\"a}t Bonn, Nussallee 12, 53115, Bonn, Germany}
\date{\today}
%%%%%%%%%%%%%%%%
\begin{abstract}
  We investigate the electronic structure at the surface of the correlated oxide \ch{Ca3Ru2O7}, a low-symmetry ruthenate oxide which hosts an unconventional polar-metal phase. From a combination of angle-resolved photoemission spectroscopy and scanning tunneling spectroscopy measurements, we demonstrate that the surface hosts an insulating phase, a distinct departure from metallicity within the bulk. Utilizing quantitative low-energy electron diffraction in conjunction with electronic structure calculations, we show how this results from a combined surface structure relaxation and the impact of marked electronic correlations in this system. Our findings highlight the proximity of \ch{Ca3Ru2O7} to an insulating metallic state, and illustrate how subtle structural distortions can control its emergent electronic phases.
\end{abstract}
%%%%%%%%%%%%%%
\maketitle
%%%%%%%%%%
In materials where the electron-electron interactions are strong, numerous nearly degenerate ground states can arise from the complex interplay of various degrees of freedom, including lattice, orbital, charge, and spin \cite{Dagotto2005}. In these so-called strongly correlated electron materials, a subtle perturbation to any of these can result in a significant change in the material's physical properties. Consequently, strongly correlated materials exhibit diverse phase diagrams that encompass exotic many-body quantum phases, including strange metallicity, unconventional superconductivity, and Mott insulators \cite{Tokura2017,Phillips2022}. The \ac{RP} ruthenates are an ideal materials class in which to explore this physics. They host a wide range of ground states which can be accessed via only rather minor modification of the crystal structure and composition, including unconventional superconductivity, magnetism and correlated insulating states \cite{Maeno1994,Lester2015,Fobes2007,Nakatsuji1997}.\par
\begin{figure}[h!]
  \centering
  \includegraphics[width=\columnwidth]{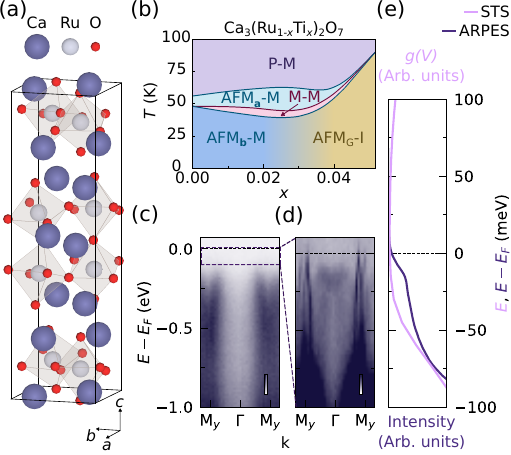}
  \caption{(a) Unit cell of bulk \ch{Ca3Ru2O7} with lattice vectors labeled.
    (b) Magnetic and electronic phase diagram of \ch{Ca3(Ru_{1-$x$}Ti_{$x$})2O7}.
    P: Paramagnetic, -M: Metallic phase, -I: Insulating phase, AFM$_\mathrm{\emph{a}}$: Antiferromagnetic order with spins orientated along the \emph{a}-axis, AFM$_\mathrm{\emph{b}}$: Antiferromagnetic order with spins orientated along the \emph{b}-axis, M: Mixed magnetic phase and AFM$_\mathrm{\emph{G}}$: Antiferromagnetic type $G$ order, with spins canted along the \emph{c}-axis. Data taken from Refs. \cite{Peng2013,Peng2016}. (c) and (d) ARPES measurements ($h\nu = 33$ eV, linear horizontal polarization, $T = 18$~K) taken along the $\mathrm{M}_y-\Gamma-\mathrm{M}_y$ direction. (e) Tunneling spectrum for the surface of \ch{Ca3Ru2O7} ($I_{\mathrm{set}} = 200$ pA, $V_{\mathrm{set}} = -200$ mV, $T = 4.2$~K) (pink line) and angle-integrated photoemission intensity extracted from panel d (purple line).}
  \label{fig:figure_1_main}
\end{figure}

In particular, \ch{Ca3Ru2O7}, the bilayer member of the \ac{RP} series of calcium ruthenates, displays competing electronic and magnetic orders, and appears to lie on the verge of multiple electronic and magnetic instabilities \cite{Kikugawa2007,Bao2008,Sokolov2019}. Upon cooling from a `bad-metal'-like metallic state at room temperature, it undergoes a sequence of magnetic and structural phase transitions. A N\'{e}el ordering occurs to an \ac{AFM} phase at $T_\mathrm{N}\approx56$~K, where the spins orient along the crystallographic \emph{a}-axis, aligning ferromagnetically within a bilayer but antiferromagnetically between neighboring bilayers (\ac{AFM}$_\mathrm{\emph{a}}$). At $T_s\approx48$~K, the spins reorient to point along the \emph{b}-axis (\ac{AFM}$_\mathrm{\emph{b}}$), coupled with an isostructural phase transition which leads to a large jump in the measured resistivity~\cite{McCall2003,Cao1997,Cao2004,Ohmichi2004}. While there has been debate regarding the nature of the ground state, the observation of quantum oscillations unambiguously points to a low carrier-density semi-metallic state~\cite{Kikugawa2010}. However, a marginal doping of the \ch{Ru} sites with non-magnetic \ch{Ti} is sufficient to drive a transition an insulating state at $\approx 4\%$ \ch{Ti} doping \cite{Ke2011,Tsuda2013}, as shown in figure \ref{fig:figure_1_main}b.\par
%%%%%%%%%%%%%%
%%%%%%%%%%%%
%%%%%%%%%%%%%%%
\begin{figure*}[ht]
  \centering
  \includegraphics[width=\textwidth]{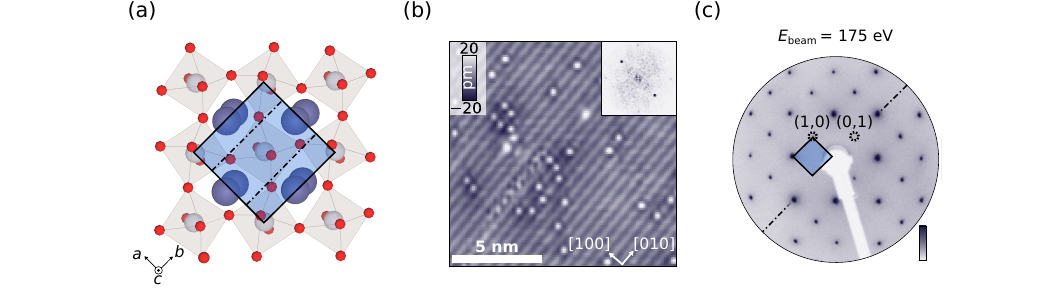}
  \caption{(a) Top-down view of the two-dimensional bulk unit cell of \ch{Ca3Ru2O7}.
    (b) Low-temperature STM topography of the \ch{Ca3Ru2O7} surface ($I = 100$ pA, $V = -2$ V, $T = 4.2$ K). Inset shows the Fourier transform of the topography. The image was low-pass filtered to remove high‑frequency noise. (c) LEED pattern of \ch{Ca3Ru2O7} measured at $T = 10$ K and an incident electron beam energy, $E_{\mathrm{beam}} = 175$ eV. The locations of the present (1,0) and extinct (0,1) LEED spots have been highlighted by the white circles. The bulk unit cell is highlighted by the blue square, and the locations of the glide line is marked by the dot-dashed lines in panels (a) and (b).}
  \label{fig:figure_2_main}
\end{figure*}
%%%%%%%%%%%%%
\newpage
This change in the electronic ground state of \ch{Ca3Ru2O7} is accompanied by a change in the low-temperature magnetic structure. Increasing the \ch{Ti} concentration, the material transitions from the \ac{AFM}$_\mathrm{\emph{b}}$ phase to an \ac{AFM}$_\mathrm{G}$ phase, with $G$-type \ac{AFM} order and spins canted along the \emph{c}-axis \cite{Peng2016}. Interestingly, the \ch{Ti}-doping into the crystal lattice drives a change in the rotation and tilt of the \ch{RuO6} octahedra \cite{Peng2013}, highlighting the sensitivity of the physical properties of \ch{Ca3Ru2O7} to subtle distortions of its constituent \ch{RuO6} octahedra.\par
This sensitivity offers rich opportunities to explore different correlated phases. Within prior studies, pressure and strain have been utilized to alter the structure of the \ch{RuO6} octahedra, and hence, the magnetic and electronic properties of \ch{Ca3Ru2O7} \cite{Karpus2006,Dashwood2023}. The local structure can also, however, be different at the surface of the crystal to the bulk~\cite{Matzdorf2002,Hu2010}, providing an ultra-clean route to track the impact of subtle structural distortions on the electronic states and phases that form. In this letter we report on the impact of such surface structural distortions in \ch{Ca3Ru2O7}, demonstrating how they not only tune the electronic structure but, in fact, also allow for a distinct, correlated insulating ground state to be stabilized.\par
Single crystal samples of \ch{Ca3Ru2O7} were grown by the floating zone method \cite{Kikugawa2021}. Both antiphase and polar domains are typically present in growths of \ch{Ca3Ru2O7}. The spatial extent of antiphase domains is such that we could select mono-domain samples for \ac{ARPES} using polarized light microscopy \cite{Markovic2020} or measure diffracted intensity from a single antiphase domain in \ac{LEED} (see supplementary figure 1). In comparison, polar domains are significantly smaller \cite{Lei2018}. Consequently, our \ac{LEED} measurements average over an approximately equal population of polar domains (see supplementary figure 2).\par
All samples were cleaved {\it in situ} before the measurements. For \ch{Ca3Ru2O7}, the intra-bilayer coupling is much stronger than the inter-bilayer coupling \cite{Yoshida2004}. This means the crystal cleaves between bilayers, exposing a \ch{CaO2} terminated surface \cite{Halwidl2017}. \ac{ARPES} measurements were performed on a freshly-cleaved surface using the BLOCH beamline of the Max-IV synchrotron. The samples were measured at a temperature of $T \approx 18$ K using 33 eV synchrotron light. \Ac{STS} measurements were performed using a home-built low temperature \acp{STM} operating in a vector magnet \cite{trainer_cryogenic_2017}. The measurements were performed at $4.2$ K unless stated otherwise. The bias voltage was applied to the sample with the tip on virtual ground. The differential conductance was recorded using a standard lock-in technique, adding an AC modulation to the sample bias. Samples for \ac{STM} are cleaved using a cleaving stage mounted to the $20$ K-plate of the insert. \ac{LEED} measurements were performed using an OCI \ac{LEED} BDL800IR optics mounted onto a \acl{UHV} system. Measurements were performed at a temperature of $T \approx 10$ K. Images of the \ac{LEED} pattern were acquired using a CCD camera, and taken in steps of 1 eV between 150 and 500 eV. Experimental $I(V,\mathbf{Q})$ spectra were extracted from the \ac{LEED} images using Python code before being analyzed using the CLEED package \cite{Held1996,Held2025}. To estimate the uncertainties associated with our \ac{LEED} analysis, we measured and fit two different samples of \ch{Ca3Ru2O7}. The displacements presented below are the mean best-fit displacement, and uncertainties are one standard deviation from the mean.\par
%%%%%%%%%%%%%%%
\begin{figure}[ht]
  \centering
  \includegraphics[width=\columnwidth]{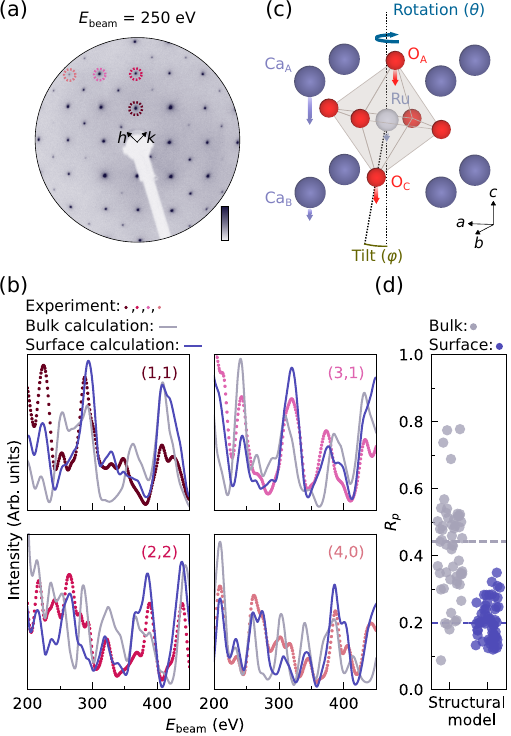}
  \caption{(a) \ac{LEED} image recorded at an incidence electron beam energy, $E_{\mathrm{beam}}$, of 250 eV. Locations of the (1,1), (2,2), (3,1) and (4,0) \ac{LEED} spots are highlighted by colored circles. (b) Four selected experimental (colored points) $I(V,\mathbf{Q})$ spectra of \ch{Ca3Ru2O7}. Calculated $I(V,\mathbf{Q})$ spectra correspond to the best-fit surface structure (purple lines) or the bulk structure taken from Ref. \cite{Yoshida2005} (gray lines). (c) The out-of-plane and in-plane displacements at the surface of \ch{Ca3Ru2O7}. (d) $R_p$ values for all the $I(V,\mathbf{Q})$ spectra analyzed and for two different structural models: the best-fit surface structure (purple points) and the bulk structure from Ref. \cite{Yoshida2005} (gray points). The total $R_p$ value for each model is indicated by the dashed line and points are offset horizontally for clarity.}
  \label{fig:figure_3_main}
\end{figure}
%%%%%%%%%%%%%
To model the surface electronic structure of \ch{Ca3Ru2O7}, \ac{DFT} calculations were performed using the \ac{VASP} code \cite{Kresse1993,Kresse1996,Kresse1996a} and local density approximation exchange-correlation functional. Both \ac{SOC} and magnetic order along the \emph{b}-axis were included within the calculations and correlations were included by adjusting the $U$ value of the \ch{Ru} atoms within the Dudarev approximation \cite{Dudarev1998}. To model the effect of structural distortions at the surface while maintaining acceptable computational cost, calculations were performed for a free-standing bilayer of \ch{Ca3Ru2O7}, with a 15 \r{A} gap to the next bilayer. A \textbf{k}-grid of 8$\times$8$\times$1 was used for self-consistent calculations, and a finer \textbf{k}-grid of 15$\times$15$\times$1 was for \ac{DOS} calculations, both with a plane wave energy cut-off of 600 eV.\par
In figure \ref{fig:figure_1_main}c, we show our \ac{ARPES} measurements over a broad energy range below the Fermi level. Consistent with prior studies, we observe a significant spectral weight in the deep-lying valence band, originating from \ch{Ru} $t_{2g}$ states \cite{Baumberger2006,Markovic2020,Horio2021}, but with broad linewidths that point to rather incoherent states. Close to the Fermi level, this spectral weight becomes strongly suppressed. Over a narrow energy range below the Fermi level, we observe a vanishing spectral weight. Nonetheless, with an enhanced contrast, we can observe sharp but extremely weak states close to and crossing the Fermi level (figure \ref{fig:figure_1_main}d), before their spectral intensity becomes dominated by the incoherent states beyond $\sim 100$ meV. The finite albeit small \ac{DOS} at the Fermi level, figure \ref{fig:figure_1_main}e, is consistent with the ground state of \ch{Ca3Ru2O7} as being a low carrier-density metal, as known from bulk studies~\cite{Kikugawa2010}. However, within this picture, the almost vanishingly-small coherent spectral weight at $E_\mathrm{F}$ as compared to the incoherent weight at higher binding energies would suggest an extremely small quasiparticle residue.\par
To better understand the electronic structure, we have therefore performed \ac{STS} measurements, displayed also in figure \ref{fig:figure_1_main}e.
These data, taken on the same batch of crystals as the \ac{ARPES} measurements, display a full gap of $\sim\!150$ meV asymmetrically distributed around the Fermi level. This indicates an insulating state, seemingly at odds with our photoemission measurements. We will show below that this is due to a distinct surface and bulk electronic structure, and different depths of the two techniques.\par
First, we consider the geometry of the \ch{Ca3Ru2O7}~(001) surface, as shown in figure \ref{fig:figure_2_main}a. In figure \ref{fig:figure_2_main}b, we present a topographic image of this \ch{CaO}-terminated surface. Prominent within the topography are stripes. These stripes represent channels with a decreased (darker) or increased (lighter) density of oxygen atoms \cite{Mayr2019}, resulting from the tilting of the \ch{RuO6} octahedra, as illustrated in figure \ref{fig:figure_2_main}a. As the tilting of the \ch{RuO6} octahedra is unidirectional in \ch{Ca3Ru2O7}, we observe stripes only along the $[010]$ crystallographic direction, and thus within the Fourier transform of the topography (inset in figure \ref{fig:figure_2_main}b) the (1,0) and (-1,0) Bragg peaks are present whilst the (0,1) and (0,-1) Bragg peaks are extinct. The unidirectional tilt of the \ch{RuO6} octahedra is also observed in our \ac{LEED} analysis of a pristine (001) surface of \ch{Ca3Ru2O7}, figure \ref{fig:figure_2_main}c. A tilting of the \ch{RuO6} octahedra in \ac{RP} crystals destroys one glide line in the two-dimensional unit cell. If the octahedral tilt is along a single axis, only a single glide line symmetry is broken whilst the orthogonal glide line is preserved. This is illustrated by the dot-dashed lines in the top-down view of the two-dimensional unit cell in figure \ref{fig:figure_2_main}a. Therefore, in our \ac{LEED} measurements, we observe spots along a single direction, the [100] direction, whilst those along the orthogonal [010] direction are extinct.\par
Together, these measurements suggest a surface symmetry consistent with the one expected from the bulk crystal structure. In fact, at all incident electron beam energies measured, we observe only \ac{LEED} spots corresponding to the bulk-terminated unit cell, indicating that \ch{Ca3Ru2O7} does not undergo a structural reconstruction at its surface. Surface relaxation can still occur, however, resulting in changes in \ch{RuO6} octahedral rotation, tilt angle or vertical displacements of atoms from their bulk positions. Indeed, the symmetry-preserving structural phase transition in the bulk indicates a strong coupling of electronic, magnetic and intra-unit-cell structural degrees of freedom in this system, necessitating searching for structural changes beyond simple reconstruction effects. This motivates quantitative \ac{LEED} measurements, recording and modeling the intensity of the \ac{LEED} spots as a function of energy in order to extract quantitative information about atomic displacements of the surface atoms.\par
%%%%%%%%%%%%%%%
\begin{table}[ht]
  \centering
  \begin{tabular}{cc}
    \toprule
    & \ac{LEED} $I(V,\mathbf{Q})$ \\
    \midrule
    $\theta_\mathrm{surface}$ & $\left(13 \pm 0.1\right)^\circ$ \\
    $\phi_\mathrm{surface}$ & $\left(13 \pm 0.5\right)^\circ$ \\
    $\Delta_{c}$ \ch{Ca_A} & $\left(-12.5 \pm 0.8\right)$ pm \\
    $\Delta_{c}$ \ch{O_A} & $\left(-5.7 \pm 0.3\right)$ pm \\
    $\Delta_{c}$ \ch{Ru} & $\left(-2.5 \pm 0.4\right)$ pm \\
    $\Delta_{c}$ \ch{Ca_B}& $\left(-3.3 \pm 0.3\right)$ pm \\
    $\Delta_{c}$ \ch{O_C} & $\left(-6.4 \pm 0.9\right)$ pm \\
    \bottomrule
  \end{tabular}
  \caption{Out-of-plane and in-plane displacements at the surface of \ch{Ca3Ru2O7}. Displacements are the mean of two samples of \ch{Ca3Ru2O7} measured and analyzed, while the uncertainties are reported as plus-minus one standard deviation from the mean. The direction of the vertical displacements and their relative magnitude are displayed in figure \ref{fig:figure_3_main}c. We note that by careful fitting of structural parameters and vibrational amplitudes we were able to achieve very low $R_p$ factors for both samples analyzed and a very high degree of precision.}
  \label{tab:LEED IV}
\end{table}
%%%%%%%%%%%%%
We show such $I(V,\mathbf{Q})$ spectra for four selected \ac{LEED} spots (highlighted in figure \ref{fig:figure_3_main}a), in figure \ref{fig:figure_3_main}b (with all the $I(V,\mathbf{Q})$ spectra shown in supplementary figure 3). Variations in the intensity of these $I(V,\mathbf{Q})$ spectra encode structural information concerning atoms within the surface layer of \ch{Ca3Ru2O7}. To elucidate these atomic coordinates, we performed a fitting to the experimental $I(V,\mathbf{Q})$ spectra, considering only structures compatible with the symmetry of the diffraction pattern seen in the \ac{LEED} pattern, figure \ref{fig:figure_2_main}c. Figure \ref{fig:figure_3_main}c highlights the key structural parameters extracted from our analysis, which are are presented in detail in table \ref{tab:LEED IV}.\par
The results of this \ac{LEED} $I(V,\mathbf{Q})$ analysis indicate a surface structure of \ch{Ca3Ru2O7} which is markedly changed from the bulk-like structure, despite the lack of a symmetry-breaking surface reconstruction. Specifically, we find \ch{RuO6} octahedra at the surface which are slightly ($\approx 2^{\circ}$) more rotated than those within the bulk crystal ($\theta_{\mathrm{bulk}} \sim 10.9^{\circ}$) \cite{Yoshida2005}, while their tilt angle ($\phi$) is essentially unaffected by the creation of a surface. Importantly, we find notable vertical displacement of surface atoms from their bulk positions. In particular, a significant vertical displacement of both \ch{Ca} atoms indicates an asymmetric uniaxial compression at the surface of \ch{Ca3Ru2O7}. Similarly, we find that both apical \ch{O} atoms displace from their bulk positions towards the underlying bulk, leading to a rigid shift of the surface \ch{RuO6} octahedra away from the vacuum interface, while the \ch{Ru} atoms undergo a vertical displacement which modifies the local symmetry of the surface octahedra, breaking the inversion symmetry of the surface bilayer. Due to destructive and constructive interference of the low-energy electron beam diffracted by the lattice, \ac{LEED} $I(V,\mathbf{Q})$ is highly sensitive to such atomic displacements, and we show in figure~\ref{fig:figure_3_main}b that the experimental and fitted $I(V,\mathbf{Q})$ spectra for such a surface are markedly different from one with a bulk-like truncation. Including such structural relaxation leads to a marked reduction of Pendry’s R-factor ($R_p$) \cite{Pendry1980} for almost all of the extracted \ac{LEED} spots (figure \ref{fig:figure_3_main}d). The global $R_p$ factor is reduced from the rather large value of $R_p^{\mathrm{bulk}} \approx 0.44$ to a best-fit value of $R_p^{\mathrm{surface}} \approx 0.20$.\par
%%%%%%%%%%%%%%%%%%
\begin{figure}[ht]
  \centering
  \includegraphics[width=\columnwidth]{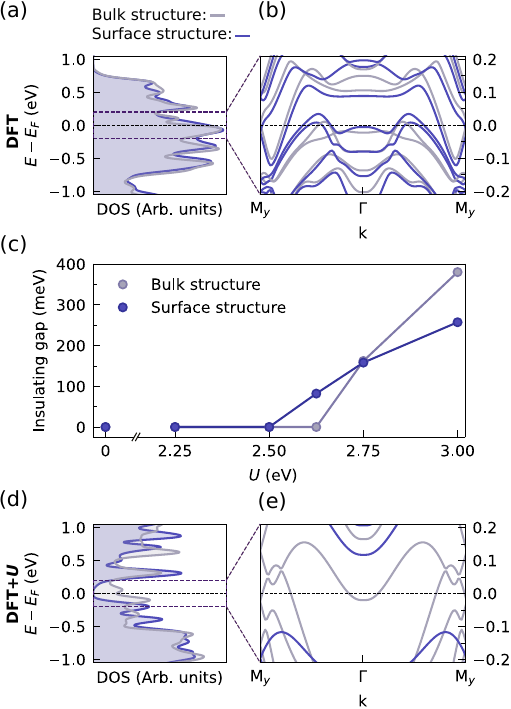}
  \caption{(a) and (b) \ac{DOS} and electronic band structure calculations for the bulk (gray lines) and surface structure (purple lines) of \ch{Ca3Ru2O7} for $U = 0$ eV. (c) Magnitude of the insulating gap around the Fermi level as a function of $U$ value, calculated for: the bulk crystal structure (gray circles) and the surface structure (purple circles). (d) and (e) \ac{DOS} and electronic band structure calculations for the bulk (gray lines) and surface structure (purple lines) of \ch{Ca3Ru2O7} for $U = 2.625$ eV. All calculations also include \ac{SOC} and magnetic order along the \emph{b}-axis.}
  \label{fig:figure_4_main}
\end{figure}
%%%%%%%%%%%%%
To investigate the impact of these atomic displacements on the surface electronic structure, we have performed a series of \ac{DFT} calculations. In figure \ref{fig:figure_4_main}a and b \ac{DFT} band structure calculations are shown for a free-standing bilayer of \ch{Ca3Ru2O7} within the low-temperature magnetic phase, \ac{AFM}$_\mathrm{\emph{b}}$. The crystal structure of the bilayer was fixed to either the bulk structure (gray) or surface structure (purple) from our \ac{LEED} $I(V,\mathbf{Q})$ analysis.\par
We find that the surface relaxation alone does not cause a significant change in the electronic structure within a normal DFT calculation. Despite an increased rotation angle of the surface \ch{RuO6} octahedra, there is no substantial narrowing of the bandwidth. However, when including electronic correlations through a local Coulomb repulsion $U$, we find a marked difference for the electronic structures. As shown in figure \ref{fig:figure_4_main}c, a band gap opens at the Fermi level with increasing $U$, pointing to a correlation-driven \ac{MIT}. While susceptibility to such a correlated-insulating state is not surprising in and of itself, intriguingly our calculations find that the surface relaxation causes this transition to occur at a different (slightly lower) $U$ value than in the bulk. For a narrow range of $U$ values ($\sim 2.5$ to $2.75$ eV), we calculate that the bulk crystal is metallic, while for the surface structure we find a correlated insulator, as shown for $U = 2.625$ eV in figure \ref{fig:figure_4_main}d and e.\par
These \ac{DFT}$+U$ calculations provide a natural explanation for the seemingly contradictory results introduced in figure \ref{fig:figure_1_main}d. While both \ac{ARPES} and \ac{STS} are surface-sensitive probes, they probe the electronic structure with different effective information depth. Within \ac{ARPES} measurements, where the surface sensitivity is determined by the mean free path of the escaping photoelectrons, there is a finite contribution to the signal from the bulk-like layers immediately below the surface. In contrast, \ac{STS} probes the tails of the wavefunctions of the surface electrons, and is thus practically only sensitive to the surface-most layer. Consequently, within our \ac{ARPES} measurements in figure \ref{fig:figure_1_main}c, we observe both the surface and (with lower intensity) bulk-like electronic states. \Ac{STM}, however, probes only an insulating surface layer. We thus assign the metallic bands with weak spectral weight visible in \ac{ARPES} to be those of the bulk crystal, consistent with prior transport studies which reveal metallicity in the bulk crystal \cite{Yoshida2004,Kikugawa2010}. Contrastingly, we attribute the incoherent states located at energies $~>~100$ meV away from the Fermi level, which carry the dominant spectral weight in \ac{ARPES} and which will dominate the gap edges in STS, as the lower and upper Hubbard bands of a Mott-like correlated insulating phase. We thus conclude that, while the subtle structural distortions of the surface layer are insufficient to significantly modify the electronic structure on their own, they facilitate the formation of a correlation driven insulating state.\par
In summary, via a combined approach of precise surface structural and spectroscopic measurements, accompanied by electronic structure calculations, we have found that a structural relaxation at the surface of \ch{Ca3Ru2O7} facilitates the formation of a correlation-driven insulating phase. This allows us to reconcile seemingly conflicting spectroscopic measurements, and reveals the striking role that surface relaxations can have, even when they do not lead to a symmetry-lowering reconstruction. Our calculations highlight that this insulating phase is rather fragile, being on the verge of metallicity. This raises an opportunity to drive a phase transition into a metallic state with moderate external stimuli, for example, via the application of strain or electric field. Furthermore, the subtle structural relaxations at the surface of \ch{Ca3Ru2O7} and the narrow $U$ window within which we calculate the surface to be insulating while the bulk remains metallic indicate that the bulk material is also on the verge of a \ac{MIT}, raising the possibility of driving the bulk through a phase transition.\par
%%%%%%%%%%%%%%%%%%%%%%%%%%%%%
\emph{Acknowledgments:} We thank L. Rhodes, I. Markovi\'{c}, C. Nicklin, and H. Hussain  for useful discussions. We gratefully acknowledge support from the Leverhulme Trust via Grant Nos.~RL-2016-006 and RPG-2022-315. D. H. gratefully acknowledges studentship support from Diamond Light Source Ltd. A.Z. and I.M. gratefully acknowledge studentship support from the International Max-Planck Research School for Chemistry and Physics of Quantum Materials. We acknowledge the MAX IV Laboratory for beamtime on the BLOCH beamline under proposal 20220162. Research conducted at MAX IV, a Swedish national user facility, is supported by Vetenskapsrådet (Swedish Research Council, VR) under contract 2018-07152, Vinnova (Swedish Governmental Agency for Innovation Systems) under contract 2018-04969 and Formas under contract 2019-02496. This work is supported by the KAKENHI Grants-in-Aid for Scientific Research (Grant No. 24K01461) from the Japan Society for the Promotion of Science (JSPS). This work used the Cirrus UK National Tier-2 HPC Service at EPCC (http://www.cirrus.ac.uk) funded by The University of Edinburgh, the Edinburgh and South East Scotland City Region Deal, and UKRI via EPSRC. The research data supporting this publication can be accessed at ref.~\onlinecite{datadoi}.
%%%%%%%%%%%%%%%%%%%%%%%%%
\bibliography{References}
%%%%%%%%%%%%%%%%%%%%%%%%

\clearpage
\onecolumngrid
\setcounter{figure}{0}
\setcounter{table}{0}
\setcounter{section}{0}

\renewcommand{\figurename}{{Supplementary Fig.}}
%%%%%%%%%%%%%%%%%%%%%%%%%%%%%%%%%%%%%%%%%%%%%%%%

%%%%%%%%%%%%%%%%%%%%%%%%%%%%%%%%%%%%%%%%%%%%%%%%%%%%%%%%%%%%%%
\begin{center}
\huge{Supplementary Information: A correlated insulator at the surface of the polar metal \ch{Ca3Ru2O7}}
\end{center}
%%%%%%%%%%%%%%%%%%%%%%%%%%%%%%%%%%%%%%%%%%%%%%%%%%%%%%%%%%%%%%
%%%%%%%%%%%
%%%%%%%%%%%%%%%%%%%%%%%%%%%%
\section*{Antiphase domains}
%%%%%%%%%%%%%%%%%%%%%%%%%%%%
%%%%%%%%%%%%%%
\begin{figure}[ht]
    \begin{center}
    \includegraphics[width=\columnwidth]{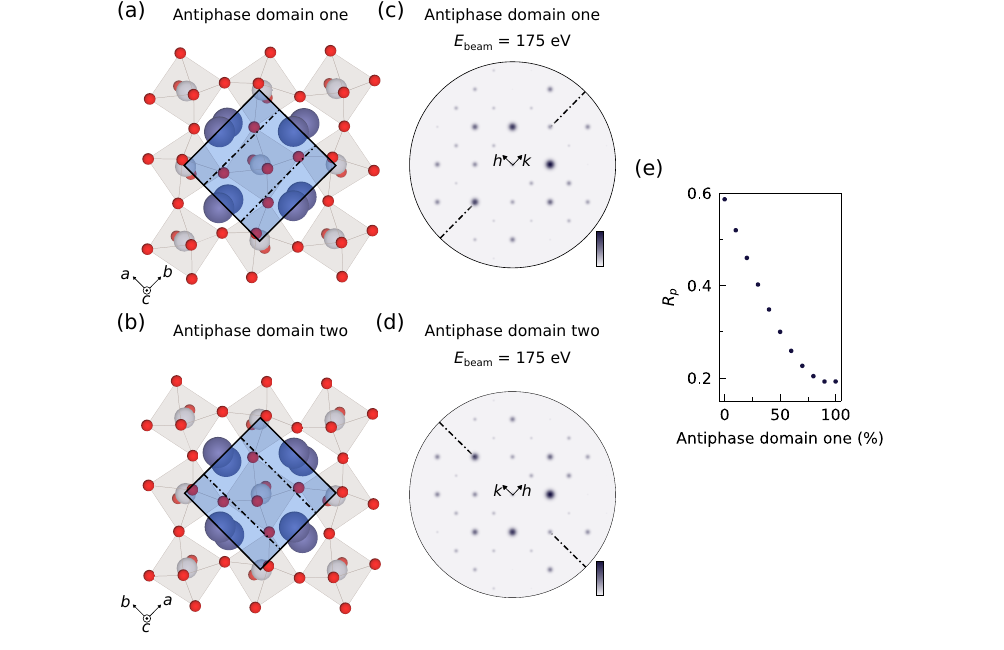}
    \end{center}
    \caption{(a) and (b) Top view of the \ch{Ca3Ru2O7} (001) surface and the two antiphase domains. Between antiphase domains the direction of the preserved glide line (dot-dashed line) is rotated by 90$^\circ$. (c) and (d) Simulated LEED patterns at an incident electron beam energy, $E_\mathrm{beam}$, of 175 eV for the two antiphase domains. As the direction of the preserved glide line is rotated by 90$^\circ$ between antiphase domains, the direction of the extinct LEED spots is also rotated by 90$^\circ$ between antiphase domains. (e) $R_p$ factor as a function of antiphase domain occupancy within the calculated $I(V,\mathbf{Q})$ spectra. The $R_p$ value has a minimum close to 100$\%$ occupancy, suggesting only a single antiphase domain is present within the region of the probing LEED spot. This is consistent with the observation of only a single extinct glide line within our LEED measurements.}
    \label{fig:figure_1_sup}
\end{figure}
%%%%%%%%%%%%
\newpage
%%%%%%%%%%%%%%%%%%%%%%%%
\section*{Polar domains}
%%%%%%%%%%%%%%%%%%%%%%%%
%%%%%%%%%%%%%%
\begin{figure}[ht]
   \begin{center}
    \includegraphics[width=\columnwidth]{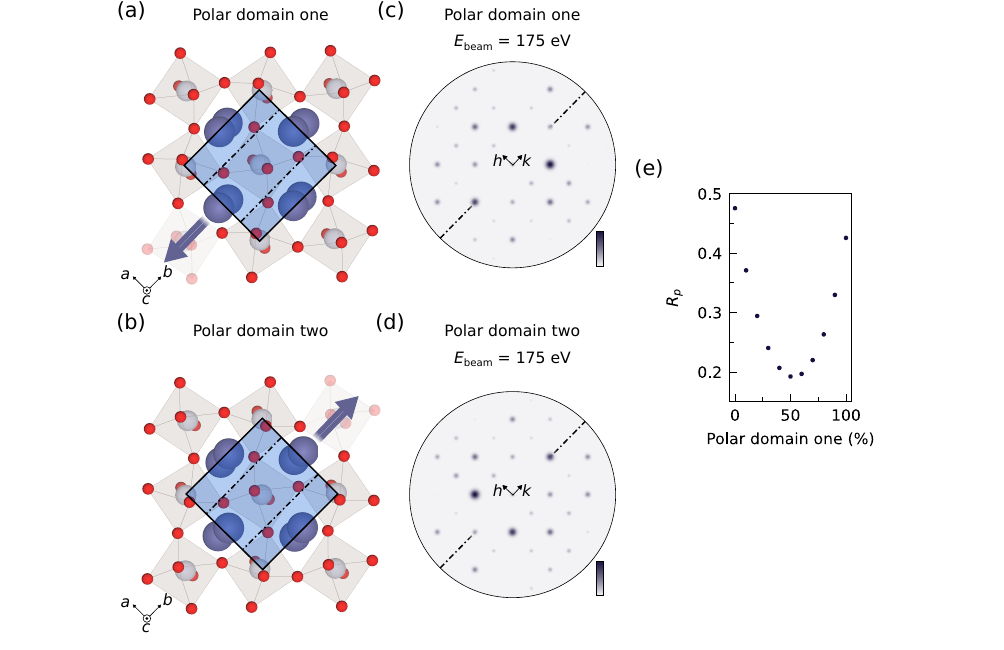}
    \end{center}
    \caption{(a) and (b) Top view of bulk \ch{Ca3Ru2O7} and the two polar domains. Between polar domains the direction of the preserved glide line (dot-dashed line) is unchanged whilst the direction of the polar displacement of \ch{Ca} atoms is reversed, as shown by the solid arrows. (c) and (d) Simulated LEED patterns at an incident electron beam energy, $E_\mathrm{beam}$, of 175 eV for the two polar domains. As the direction of the preserved glide line is unchanged between polar domains, the direction of the extinct LEED spots is also unchanged between antiphase domains. (e) $R_p$ factor as a function of polar domain occupancy within the calculated $I(V,\mathbf{Q})$ spectra. The $R_p$ value has a minimum at $\sim 50\%$ occupancy, suggesting an equal population of the two domains are present within the region of the probing LEED spot.}
    \label{fig:figure_2_sup}
\end{figure}
%%%%%%%%%%%%
\newpage
%%%%%%%%%%%%%%%%%%%%%%%%%%%%%%%%%%%%%%%%%
\section*{LEED $I(V,\mathbf{Q})$ analysis}
%%%%%%%%%%%%%%%%%%%%%%%%%%%%%%%%%%%%%%%%%
%%%%%%%%%%%%%%
\begin{figure}[h!]
    \begin{center}
    \includegraphics[width=\columnwidth]{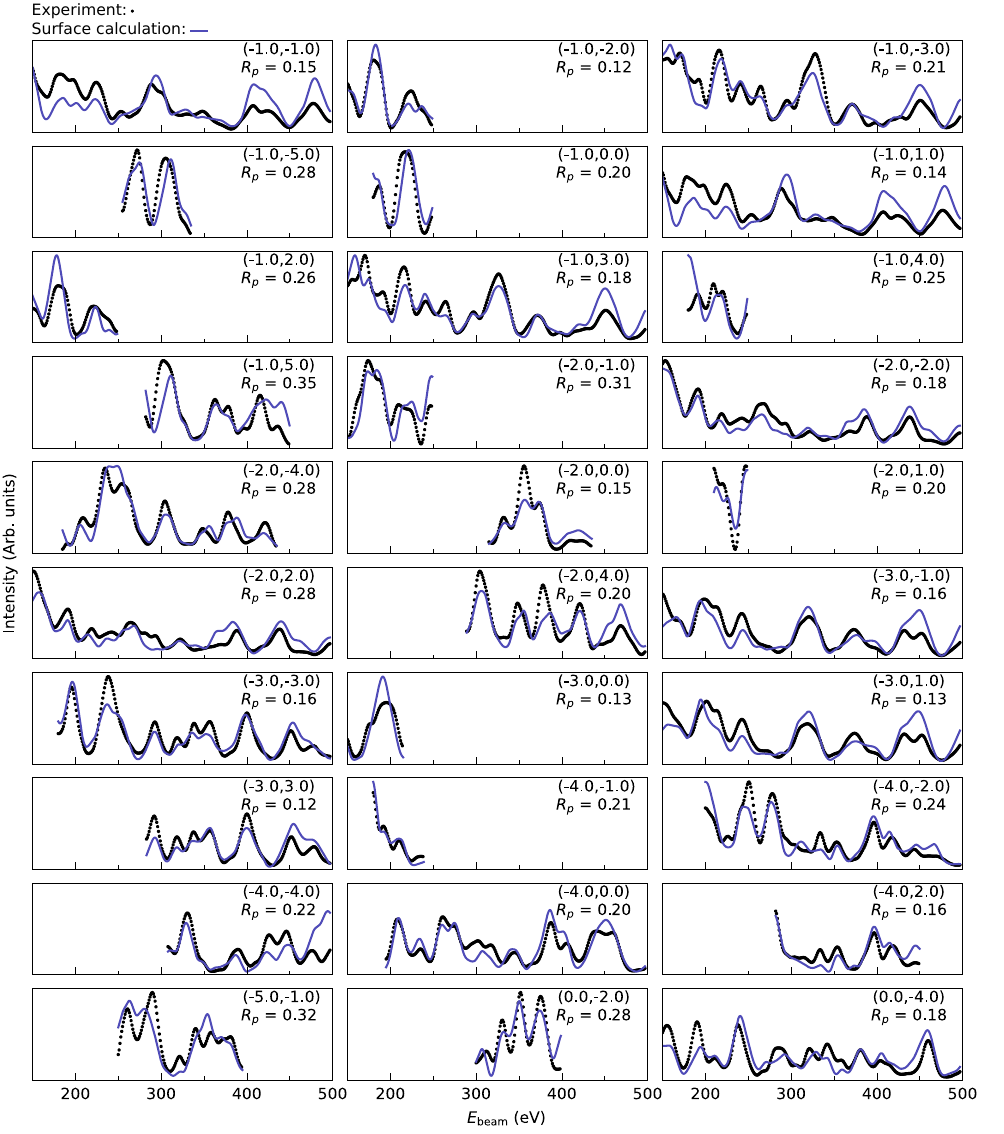}
    \end{center}
    \caption[]{Experimental (black points) $I(V,\mathbf{Q})$ spectra of \ch{Ca3Ru2O7}. Calculated $I(V,\mathbf{Q})$ spectra correspond to the best-fit surface structure (purple lines). The indices of the LEED spots are given in each panel, together with the corresponding $R_p$ values quantifying the agreement between the experimental and calculated spectra.}
    \label{fig:figure_3_sup}
\end{figure}
%%%%%%%%%%%%
\newpage
%%%%%%%%%%%%%%
\begin{figure}[h!]
   % \ContinuedFloat
    \begin{center}
    \includegraphics[width=\columnwidth]{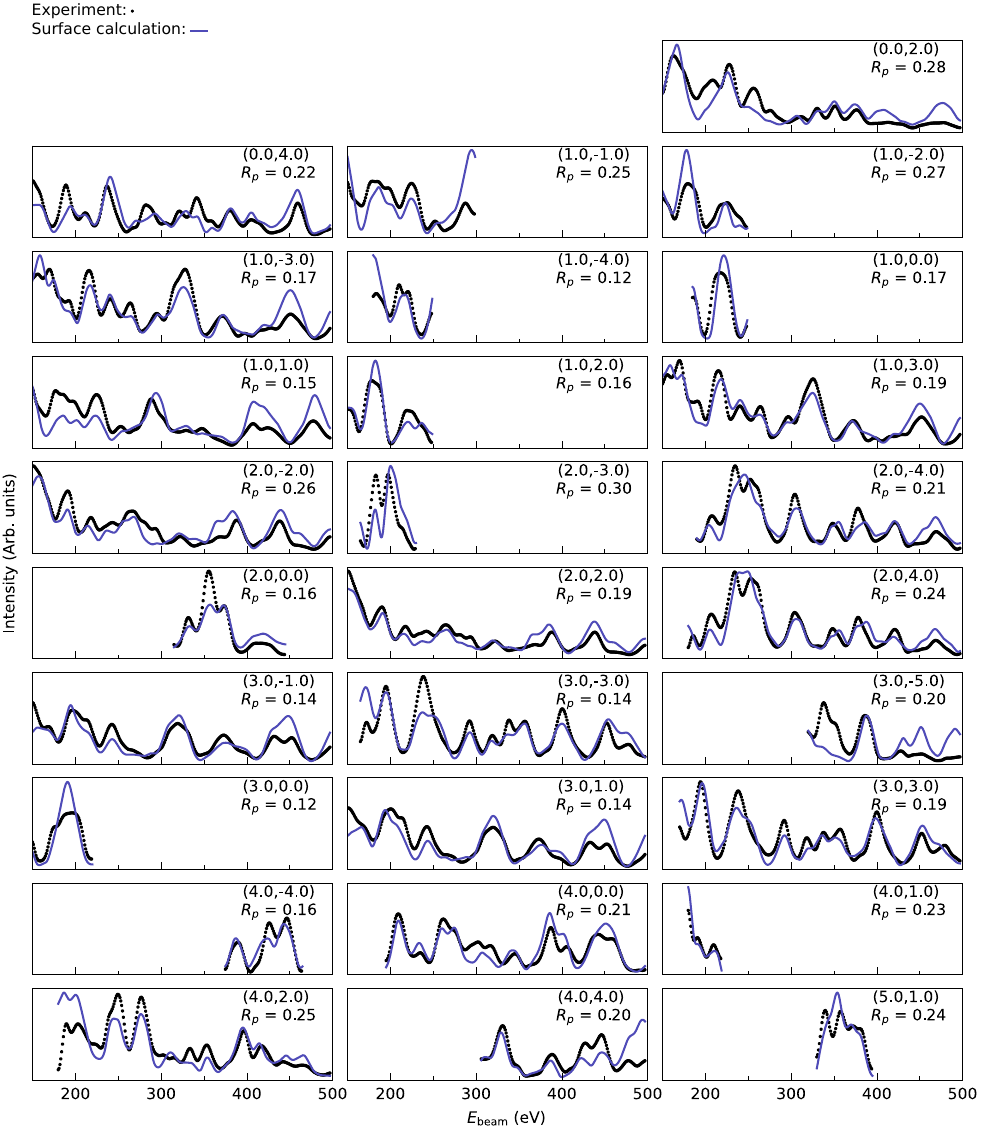}
    \end{center}
    \caption[Supplementary Fig. 3.]{(continued)}
\end{figure}
%%%%%%%%%%%%
The detailed files with the atomic coordinates resulting from the fits will be published in ref.~\onlinecite{datadoi}.
%%%%%%%%%%%%%%
%\bibliography{References}
\end{document}